\documentclass{PoS}

\usepackage{tikz}
\usetikzlibrary{trees}
\usetikzlibrary{decorations.pathmorphing}
\usetikzlibrary{decorations.markings}

\usepackage{subfigure}

\title{Landau gauge gluon vertices from Lattice QCD}

\ShortTitle{Landau gauge gluon vertices from Lattice QCD}

\author{Anthony G. Duarte, Orlando Oliveira, \speaker{Paulo J. Silva}\\
       CFisUC, Department of Physics, University of Coimbra\\
       P-3004-516 Coimbra, Portugal\\
       E-mail: \email{psilva@uc.pt}}

\abstract{In lattice QCD the computation of one-particle irreducible (1PI) Green's functions with a large number (> 2) of legs is a challenging task. Besides tuning the lattice spacing and volume to reduce finite size effects, the problems associated with the estimation of higher order moments via Monte Carlo methods and the extraction of 1PI from complete Green's functions are limitations of the method. Herein, we address these problems revisiting the calculation of the three gluon 1PI Green's function.}

\FullConference{34th annual International Symposium on Lattice Field Theory\\
         24-30 July 2016\\
         University of Southampton, UK}

\begin{document}

\section{Introduction and motivation}

In a quantum field theory, the Green's functions summarize the dynamics of 
the theory. In particular, in QCD the Green's functions provide valuable 
information about non-perturbative phenomena, like confinement and chiral 
symmetry breaking.

A $n$-point complete Green's function defined as
\begin{equation}
G^{(n)} (x_1, \dots , x_n )  = \langle 0 | T \left( \phi(x_1) \cdots \phi(x_n) \right) | 0 \rangle \, ,
\end{equation}  
may be decomposed in terms of one particle irreducible (1PI) functions 
$\Gamma^{(n)}$, which can be parame\-tri\-zed in terms of various scalar 
form factors. Although the lattice approach allows for a first principles 
determination of the complete Green's functions of QCD, in general it is
only able to compute suitable combinations of the form factors.

Here we focus on the three gluon vertex in Landau gauge, which plays a 
fundamental role in the physics of the strong interactions. Indeed, from the 
three gluon vertex it is possible to compute the strong coupling constant 
or to measure a static potential between color charges.
Furthermore, under the assumption 
that the ghost propagator remains essentially massless over all range of 
momenta, the requirement that the Dyson-Schwinger equations (DSE) are finite
implies that some of the higher order gluonic Green's functions should change
sign in the infrared region. This is the case of the three gluon vertex. Such
zero crossing has already been observed in 
continuum approaches \cite{bip, abip}, SU(2) 3d lattice simulations \cite{cucc1, cucc2}, and very recently in 
SU(3) 4d lattice simulations \cite{binosi}.\footnote{See \cite{sternbeck} for another recent lattice calculation of the three gluon vertex.}  The DSE analysis suggests that, for SU(3), the 
zero crossing should occur at a momentum scale $\sim 130 - 200$ MeV \cite{abip}.

In momentum space, the three point complete Green's function $G^{a_1 a_2 a_3}_{\mu_1 \mu_2 \mu_3} (p_1, p_2, p_3)$ is defined as
\begin{equation}
  \langle A^{a_1}_{\mu_1} (p_1) \, A^{a_2}_{\mu_2} (p_2) \, A^{a_3}_{\mu_3} (p_3) \rangle =   V \, \delta( p_1 + p_2 + p_3) ~    G^{a_1 a_2 a_3}_{\mu_1 \mu_2 \mu_3} (p_1, p_2, p_3)
\end{equation}
and, in terms of the gluon propagator 
\begin{equation}
D^{ab}_{\mu\nu}(p)=\delta^{ab} P_{\mu\nu}(p)D(p^2) \, , \, 
P_{\mu\nu}(p)=\delta_{\mu\nu} - \frac{p_\mu p_\nu}{p^2}\, ,
\end{equation}
and the 1PI vertex $\Gamma$, it reads
\begin{equation}
G^{a_1a_2a_3}_{\mu_1\mu_2\mu_3} (p_1, p_2, p_3)  =   D^{a_1b_1}_{\mu_1\nu_1}(p_1) ~ D^{a_2b_2}_{\mu_2\nu_2}(p_2) ~ D^{a_3b_3}_{\mu_3\nu_3}(p_3) 
  \Gamma^{b_1b_2b_3}_{\nu_1\nu_2\nu_3} (p_1, p_2, p_3) .
\end{equation}
The color structure of the 1PI vertex is given by
\begin{equation}
  \Gamma^{a_1 a_2 a_3}_{\mu_1 \mu_2 \mu_3} (p_1,  p_2, p_3) = f_{a_1 a_2 a_3} \Gamma_{\mu_1 \mu_2 \mu_3} (p_1, p_2, p_3) .
\end{equation}

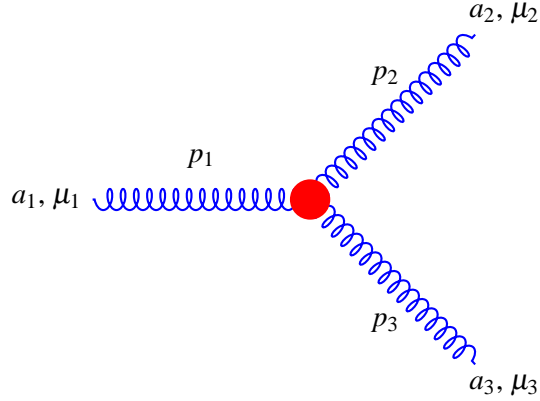
\begin{figure}[t]
\begin{center}
\tikzset{
    photon/.style={decorate, decoration={snake}, draw=red},
    electron/.style={draw=blue, postaction={decorate},
        decoration={markings,mark=at position .55 with {\arrow[draw=blue]{>}}}},
    gluon/.style={decorate, draw=blue, decoration={coil,amplitude=4pt, segment length=5pt} }
}
\begin{tikzpicture}[
        thick,
        % Set the overall layout of the tree
        level/.style={level distance=3.5cm},
        level 2/.style={sibling distance=2.6cm},
        level 3/.style={sibling distance=2cm}
    ]
    \coordinate
    child[grow=south east] {
                node {\,\, $a_3$, $\mu_3$}
                edge from parent [gluon] node [below = 8pt] {\hspace*{-2mm}$p_3$}
            };
    \coordinate
    child[grow=north east] {
                node {\,\, $a_2$, $\mu_2$}
                edge from parent [gluon] node [above = 8pt] {\hspace*{-2mm}$p_2$}
            };
    \coordinate
     child[grow=west] {
                node {$a_1$, $\mu_1$}
                edge from parent [gluon] node [above = 8pt] {$p_1$}
            };
\draw [red, ultra thick, fill] (0,0) circle [radius=0.24];
\end{tikzpicture}
\end{center}
\caption{Three gluon vertex. All momenta are incoming.}
\end{figure}

Bose symmetry requires the vertex to be symmetric under the interchange of any pair $(p_i, a_i, \mu_i)$, therefore $\Gamma_{\mu_1 \mu_2 \mu_3} (p_1, p_2, p_3)$ must be antisymmetric under the interchange of any pair $(p_i, \mu_i)$.

In the continuum, a complete description of $\Gamma_{\mu_1 \mu_2 \mu_3} (p_1, p_2, p_3)$ requires six Lorentz invariant scalar form factors. See \cite{ballchiu} for details.

\section{Lattice setup}

We have performed lattice simulations for pure SU(3) Yang-Mills theory using 
the Wilson gauge action, for $\beta=6.0$. We considered two different lattice 
volumes, $64^4$, with 2000 configurations, and $80^4$, with 279 configurations. 
All configurations have been rotated to the Landau gauge using the FFT-SD 
method \cite{davies}, which was implemented combining Chroma \cite{chroma} 
and PFFT \cite{PFFT} libraries. For the definition of the gluon field we use
\begin{equation}
a g_0 A_\mu (x + a \hat{e}_\mu/2)  = \frac{ U_\mu (x) - U^\dagger (x)}{ 2 i } 
      - \frac{ \mbox{Tr} \left[ U_\mu (x) - U^\dagger (x) \right]}{6 i },
\end{equation}
with the definition in momentum space given by
\begin{equation}
 A_\mu (\hat{p}) = \sum_x e^{- i \hat{p} (x + a \hat{e}_\mu/2) } \, A_\mu (x + a \hat{e}_\mu/2) \quad , \quad \hat{p}_\mu = \frac{2 \, \pi \, n_\mu}{a \, L_\mu}.
\end{equation}
Besides $\hat{p}_\mu$ we also use the tree-level improved momentum
\begin{equation}
   p_\mu = \frac{2}{a} \, \sin \left( \frac{a \,  \hat{p}_\mu}{2} \right)
\end{equation} 
in the description of the results.

In order to access the 1PI three gluon vertex from the lattice we 
consider the color trace
\begin{eqnarray}
&& G_{\mu_1 \mu_2 \mu_3} (p_1, p_2, p_3 ) =
 \mbox{Tr} ~ \langle A_{\mu_1} (p_1) \, A_{\mu_2} (p_2) \, A_{\mu_3} (p_3)    \rangle = \nonumber \\
&& = V \delta(p_1 + p_2 + p_3) \frac{N_c(N^2_c-1)}{4} ~ D(p^2_1) \, D(p^2_2) \, D(p^2_3) ~ \nonumber  \\
&& P_{\mu_1\nu_1}(p_1) \,  P_{\mu_2\nu_2}(p_2) \, P_{\mu_3\nu_3}(p_3) ~ \color{black}{\Gamma_{\nu_1 \nu_2 \nu_3} (p_1, p_2, p_3)} 
\end{eqnarray}
where $\langle \cdots \rangle$ means average over gauge configurations. 
Of all possible momentum configurations, in this work we only investigate 
the case with one vanishing momentum $p_2=0$. This momentum configuration 
has been used in the first lattice study of the three gluon vertex \cite{alles}.
For this kinematics
\begin{equation}
   G_{\mu_1\mu_2\mu_3} (p, 0, -p)  =    V \frac{N_c(N^2_c-1)}{4}  \left[D(p^2)\right]^2 \, D(0) \frac{\Gamma (p^2)}{3} ~ ~ p_{\mu_2} ~T_{\mu_1\mu_3} (p)
\end{equation}
and, in terms of the Ball-Chiu decomposition \cite{ballchiu}, 
$\Gamma (p^2)$ reads
\begin{equation}
  \Gamma (p^2)   =  2 \bigg[ A(p^2,p^2;0) +  p^2 \, C(p^2,p^2;0) \bigg].
\end{equation}
Here we report the form factor  $\Gamma (p^2)$ as measured from the combination
\begin{equation}
     G_{\mu \, \alpha \,\mu} (p, 0, -p) \, p_\alpha = V \frac{N_c(N^2_c-1)}{4}
   \, \left[D(p^2)\right]^2 \, D(0) ~~\Gamma (p^2) ~~ p^2.
\end{equation}

\section{The infrared region}

\begin{figure}[h]
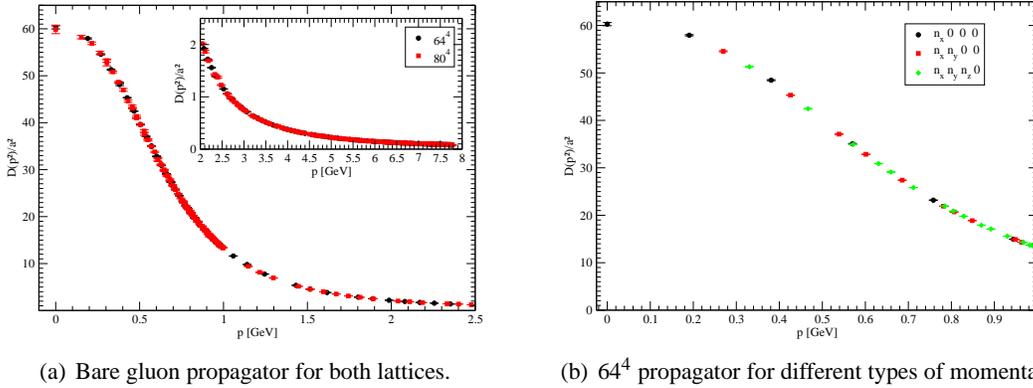
 %  figure placement: here, top, bottom, or page
\vspace{0.55cm}
   \centering
   \subfigure[Bare gluon propagator for both lattices.]{ \includegraphics[width=0.42\textwidth]{plots/Prop_64_80.eps} \label{glprop1}} \qquad
   \subfigure[$64^4$ propagator for different types of momenta.]{ \includegraphics[width=0.42\textwidth]{plots/Prop_64_XXXX.eps} \label{glprop2}}
  \caption{Bare Landau gauge gluon propagator.}
   \label{glprop}
\end{figure}

In Figure \ref{glprop1} we report the bare gluon propagator for both lattice 
ensembles described in the previous section. No clear volume effects are seen 
in the data.\footnote{See \cite{OliSi2012, dos2016} for an analysis of finite volume and finite lattice spacing effects in Landau gauge two-point correlation functions.} Moreover, in Figure \ref{glprop2} the lattice data define a 
unique curve for different types of momenta, and therefore no rotational
symmetry breaking effects are seen in the propagator. 

\begin{figure}[h]
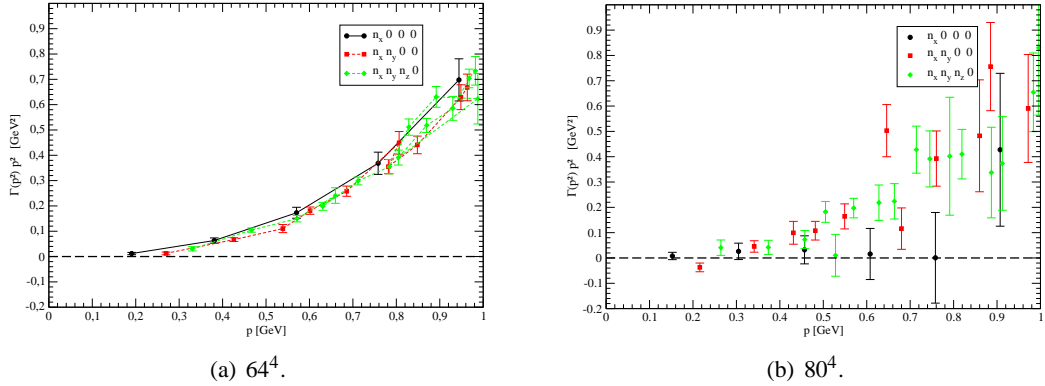
 %  figure placement: here, top, bottom, or page
\vspace{0.5cm}
   \centering
   \subfigure[$64^4$.]{ \includegraphics[width=0.42\textwidth]{plots/1PI_64_mom_noprops_XXXX.eps} } \qquad
   \subfigure[$80^4$.]{ \includegraphics[width=0.42\textwidth]{plots/1PI_80_mom_noprops_XXXX.eps} }
  \caption{Infrared $\Gamma(p^2)p^2$ for different types of momenta.}
   \label{gammamom}
\end{figure}

In Fig. \ref{gammamom} $\Gamma(p^2)$ is plotted for different types of 
momenta. 
For the $64^4$ lattice, the data for $(n000)$ momenta 
does not follow the same curve as the other types of momenta. 
Furthermore, for the $80^4$ lattice, momenta of type $(n000)$ does not 
provide any useful information about the 
form factor. As a consequence, from now on we will disregard the momenta 
of type $(n000)$ in our analysis.

\begin{figure}[t]
\vspace*{0.85cm}
\begin{center}
\includegraphics[width=0.6\textwidth]{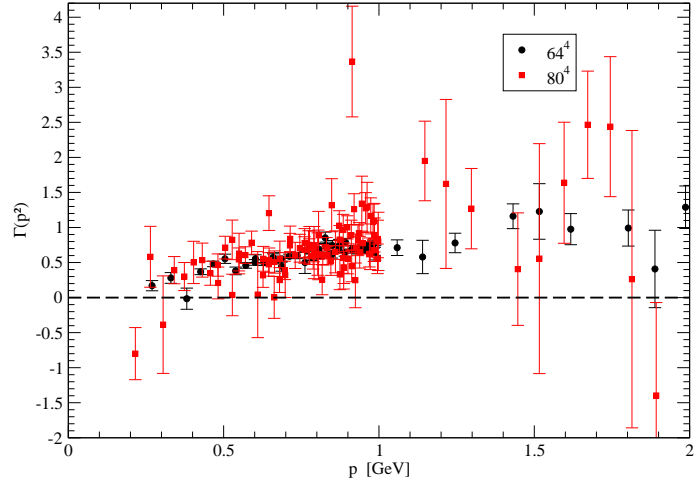}
\end{center}
\caption{$\Gamma(p^2)$ for the simulations considered in this work.}
\label{gammair}
\end{figure}

The form factor $\Gamma(p^2)$ for both data sets and for momenta up to 2 GeV 
can be seen in Fig. \ref{gammair}. Although we see larger statistical errors 
for the larger volume, the results from the two lattice ensembles are 
essentially compatible within errors. In what concerns the behaviour in 
the low momenta region, we see a negative $\Gamma (p^2)= -0.80(37)$ 
at $p = 216$ MeV for the larger lattice. This value is compatible with zero 
only within 2.2 $\sigma$. Note that for higher momenta, we have 
$\Gamma (p = 270 \mbox{ MeV}) = 0.171(73)$ from the $64^4$ volume and  
$\Gamma (p = 264 \mbox{ MeV}) = 0.58(43)$ from the $80^4$ volume. In this
sense, our data suggests that a zero crossing in $\Gamma(p^2)$ should take 
place for $p\lesssim 250$ MeV. Earlier lattice simulations reported a zero
crossing at essentially the same momentum value.

\section{The ultraviolet region}

The measurement of  $\Gamma (p^2)$ requires the computation of the ratio 
\begin{equation}
 \frac{G_{\mu \alpha \mu} (p, 0, -p) p_\alpha}{ \left[D(p^2)\right]^2 \, D(0)}.
\end{equation}
This ratio induces large statistical fluctuations at high momenta. In fact, 
if we assume gaussian error propagation for the estimation of the statistical 
error on $\Gamma (p^2)$, we obtain $\Delta \Gamma (p^2) \sim p^2 $. However, 
it is possible to measure the following combination
\begin{equation}
\Gamma_{UV} (p^2) = \left[ D(p^2) \right]^2 \, D(0) ~ \Gamma (p^2) ~ p^2 
\end{equation}
with controllable statistical errors.
Combining the predictions from one-loop renormalization group improved 
perturbation theory for $D(p^2)$ and $\Gamma (p^2)$, we get the following 
result for the behaviour of $\Gamma_{UV} (p^2)$ at high momenta:
\begin{equation}
    \Gamma_{UV} (p^2) =  \frac{Z}{p^2} \left[ \ln \frac{p^2}{\mu^2} \right]^{ \gamma^{\prime} }
\label{gammapert}
\end{equation}
where $\gamma^\prime = - 35/44$, $Z$ is a constant and $\mu$ is a 
renormalization scale.

\begin{figure}[t]
\vspace*{0.5cm}
\begin{center}
\includegraphics[width=0.6\textwidth]{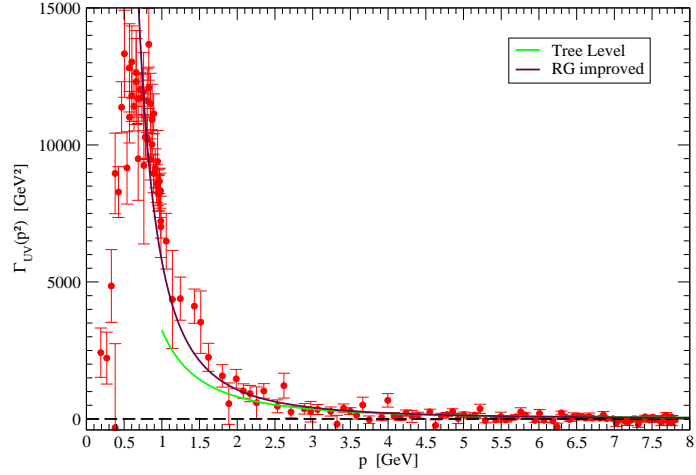}
\end{center}
\caption{Comparison of $\Gamma_{UV} (p^2)$ computed from the $64^4$ simulation with the predictions from perturbation theory.}
\label{gammauv}
\end{figure}

The results for $\Gamma_{UV} (p^2)$ can be seen in Fig. \ref{gammauv} where we also show the prediction  from tree-level perturbation theory $\Gamma_{UV} (p^2) \sim \frac{1}{p^2}$ and the prediction of Eq. (\ref{gammapert}).

\section{Conclusions}

We have computed the three gluon complete Green's function on the lattice, 
for a particular kinematical configuration ($p_2=0$), using two different 
lattice volumes, $(6.5$ fm$)^4$ and $(8.2$ fm$)^4$ for the same lattice 
spacing ($a = 0.102$ fm).
In what concerns the low momenta region, we verified that the form factor 
$\Gamma (p^2)$ exhibits a zero crossing for $p \sim 250$ MeV. 
Earlier results for 3d SU(2) \cite{cucc1, cucc2} and 4d SU(3) \cite{binosi} 
lattice simulations are in good agreement with ours.
We have also observed that, for sufficiently high momenta, the lattice 
data is compatible with the prediction of renormalisation group 
improved perturbation theory.
More details about our work can be found in \cite{dos}.

\acknowledgments

The present work was financially supported by FCT Portugal with 
reference UID/FIS/04564/\-2016. The computing time was provided by 
the Laboratory for Advanced Computing at the University of Coimbra 
\cite{lca} and by PRACE projects COIMBRALATT (DECI-9) and 
COIMBRALATT2 (DECI-12). Work of P. J. Silva partially supported by FCT 
under Contracts No. SFRH/BPD/40998/\-2007 and SFRH/BPD/109971/2015.

\end{document}